
\input phyzzx
\hsize=6.4 true in \hoffset=0.2 true in
\vsize=9.0 true in \voffset=0.2 true in

\baselineskip= 16pt
\rightline{YUMS 93--25}
\rightline{SNUTP 93--75}
\rightline{(September 1993)}
\bigskip\bigskip

\centerline{\bf  ELECTROWEAK DISCUSSION SECTION SUMMARY}
\bigskip\bigskip

\centerline{\bf C.S. Kim\foot{\rm Talk given at the 14th Interanational
Workshop on Weak Interactions and Neutrinos, July 19--25, 1993, Seoul, Korea}}
\medskip
\centerline{Department of Physics, Yonsei University, Seoul 120--749, KOREA}

\bigskip\bigskip
\centerline{\bf ABSTRACT}
\medskip


\noindent
I summarize our work of electroweak discussion section held at the 14th
International
Workshop on Weak Interactions and Neutrinos.
We discussed about a few physics topics related to electroweak interactions,
including $B$ and $K$ physics and weak flavor physics.

\hsize=6.4 true in \hoffset=0. true in
\vsize=9.0 true in \voffset=0. true in
\bigskip\bigskip

\leftline{\bf 1. INTRODUCTION}
\medskip

In the standard $SU(2) \times U(1)$ gauge theory of Glashow, Salam and
Weinberg the fermion masses and hadronic flavor changing weak transitions
have a somewhat less secure role, since they require  a prior knowledge
of the mass generation mechanism. The simplest possibility to give mass
to the fermions in the theory makes use of Yukawa interactions involving
the doublet Higgs field. These interactions give rise to the
Cabibbo--Kobayashi--Maskawa (CKM) matrix: Quarks of different flavor are
mixed in the charged weak currents by means of an unitary matrix $V$.
However, both the electromagnetic current and the weak neutral current remain
flavor diagonal. Second order weak processes such as mixing and CP--violation
are even less secure theoretically, since they can be affected by both
beyond the Standard Model virtual contributions, as well as new physics
direct contributions. Our present understanding of CP--violation is based
on the three--family Kobayashi--Maskawa model$^{1)}$ of quarks, some of whose
charged--current couplings have phases. Over the past decade, new data have
allowed one to refine our knowledge about parameters of this matrix $V$.

In Section 2, we introduce the quark mixing matrix, and a new set of quark
mass matrices is proposed based on experimental mass hierarchy. In Section 3,
we point out that directly measuring the invariant mass of final hadrons
in $B$--meson semileptonic decays offers an alternative way to select
$b \rightarrow u$ transitions that is in principle more efficient than
selecting the upper end of the lepton energy spectrum. In Section 4,
we briefly discuss about the role of vector mesons in rare kaon decays, and
about probing supergarvity models from the $b \rightarrow s \gamma$ decays.

\medskip
\leftline{\bf 2. FLAVOR DEMOCRACY AND FLAVOR GAUGE THEORY}
\medskip

It has turned out that for many fields of physics symmetry breaking
is as fundamental a feature as symmetry itself.
Whether one considers for instance
the breakdown of rotational symmetry in a ferromagnet
or the collapse of gauge symmetry in a superconductor,
it is clear that our world would not be the same if these symmetries
were respected at all temperatures.
More fundamentally,
it is the asymmetry of the electroweak vacuum
that garanties our very existence:
in the absence of a spontaneous gauge symmetry breaking mechanism,
quarks and pions would remain massless,
and nuclear matter unbound.
The hierarchical pattern of the quark masses and their mixing,
nevertheless remains an outstanding issue of the electroweak theory.
While a gauge interaction is characterized by its universal coupling constant,
the Yukawa interactions have as many coupling constants as there are fields
coupled to the Higgs boson.
There is no apparent underlying principle which governs the hierarchy of the
various Yukawa couplings,
and, as a result,
the standard model of strong and electroweak interactions
can predict neither the quark (or lepton) masses nor their mixing.

The gauge invariant Yukawa Lagrangean involving quark fields is
$$
{\cal L}_Y = - \sum_{i,j} (\bar Q'_{iL}~\Gamma^D_{ij}~d'_{jR}~\phi~+~
\bar Q'_{iL}~\Gamma^U_{ij}~u'_{jR}~\tilde \phi~+~h.c.)
\ ,
\eqno(1)
$$
where the primed quark fields are eigenstate of the $SU(2) \times U(1)$
electroweak gauge interaction.
The left-handed quarks form a doublet under the $SU(2)$ transformation,
$\bar Q'_L=(\bar u'_L,~\bar d'_L)$, and the right-handed quarks are singlets.
The indices $i$ and $j$ run over the number of fermion generations.
The Yukawa coupling matrices $\Gamma^{U,D}$ are arbitrary
and not necessarily diagonal.
The Higgs field $\phi$ is parametrized in the unitary gauge as
$$
\phi = {1\over\sqrt{2}} \pmatrix{0\cr v+H\cr}
\ .
\eqno(2)
$$
After spontaneous symmetry breaking it acquires a non-vanishing vacuum
expectation value $v$
which yields quark mass terms in the original Lagrangean
$$
{\cal L}_{mass} = -{v \over \sqrt{2}}~\sum_{i,j} (\bar d'_{iL}~\Gamma^D_{ij}~
d'_{jR}~+~\bar u'_{iL}~\Gamma^U_{ij}~u'_{jR}~+~h.c.)
\ .
\eqno(3)
$$
The quark mass matrices are defined as
$$
M^{U,D}_{ij} \equiv {v \over \sqrt{2}}~\Gamma^{U,D}_{ij}
\ .
\eqno(4)
$$
These mass matrices $M^{U,D}$ can be diagonalized
with the help of unitary matrices, $U^{U,D}_L$ and $U^{U,D}_R$,
and the electroweak eigenstates are tranformed to physical mass eigenstates
by the same unitary transformations,
$$
U^{U,D}~M^{U,D}~(U^{U,D})^{\dagger} = M^{U,D}_{diag}
\ ,
$$
$$
U^U~u'_{L,R} = u_{L,R} \qquad {\rm and} \qquad U^D~d'_{L,R} = d_{L,R}
\ .
\eqno(5)
$$
The kinetic and neutral current terms of electroweak gauge interactions
remain invariant under the unitary transformation (5).
However, the charged current terms display non-diagonal couplings
when written in terms of the physical quarks.
Defining the CKM matrix as
$$
V = U^U~U^{D\dagger}
\ ,
\eqno(6)
$$
the charged current Lagrangean becomes
$$
{\cal L}_{c.c.} \sim \sum_{i,j,k} \left[ \bar u_{iL}~\gamma^\mu~
V_{ij}~d_{jL}~W^{+}_\mu~+~\bar d_{iL}~\gamma^\mu~
V_{ij}^{\dagger}~u_{jL}~W^{-}_\mu \right]
\ .
\eqno(7)
$$

In the past, many attempts at unifying the gauge interactions of the standard
model (SM) have been made, within the framework of the Grand Unified Theories
(GUT).  These  theories yield the unification energy
$E_{\rm GUT} \geq  10^{16}$ GeV, i.e. the energy where the SM gauge couplings
are to coincide:
$$
{5 \over 3} \alpha_1 = \alpha_2 = \alpha_3~~.
\eqno(8)
$$
Here,  $\alpha_j =  g_j^2 / 4 \pi$, and $g_1,~g_2,~g_3$ are the gauge
couplings of $U(1)_Y,~SU(2)_L$, $SU(3)_C$, respectively. For the condition
(8) to be satisfied at a single point $\mu (=E_{\rm GUT})$ exactly,
supersymmetric theories (SUSY) were introduced to replace SM at
$\mu \approx M_{\rm SUSY} \approx 1$ TeV. This changed the slopes
of the functions $\alpha_j=\alpha_j (\mu)$ at $\mu \geq M_{\rm SUSY}$, and
for certain values of parameters of SUSY the three lines met at a single
point, as required by (8).
However, such an approach has several defficiencies. Due to large lower
bound for the proton decay time ($\tau_{\rm proton} \geq 5.5 \times 10^{32}$
yr), the unification energy is exceedingly large ($E_{\rm GUT} \geq 10^{16}$
GeV). This would imply that there is a large desert between $M_{\rm SUSY}$
and $E_{\rm GUT}$. Furthermore, although the initial aim of the attempts
leading to SUSY and GUT was to reduce the number of independent parameters
(gauge couplings), the overall number of free parameters was
in fact substantially increased through the inclusion of SUSY degrees of
freedom (sfermions, additional Higgses, etc.).

We believe that it is more reasonable to attempt first to reduce the number
of degrees of freedom in the sector of Higgses and Higgs-fermion interactions
(i.e. Yukawa sector), since this sector seems to be more problematic than
the gauge boson sector. Furthermore, we believe that any such attempt
should be required to lead to an overall reduction of the seemingly
independent degrees of freedom, unlike the GUT--SUSY approach. A flavor
gauge theory (FGT)$^{2)}$ is a theory having universal strengths of the
Yukawa (running) couplings at energies $E \geq \Lambda_{\rm FGT}$.
The symmetry responsible for this reduction of the number of parameters
would be ``flavor democracy'' (FD), valid possibly in certain separate
sectors of fermions. Although FGT could imply that the Higgs(es)  of SM
are elementary at low energies, it appears natural that FGT is without
Higgs and that the Higgs particles of SM are bound states of fermion pairs
$\langle \bar f f \rangle$, created through dynamical symmetry breaking
(DSB) in a transition interval $[E_{\rm trans.},~\Lambda_{\rm FGT}]$ between
SM and FGT. The idea of FD and the deviation from the exact FD at the low
energy ($E \sim 1$ GeV) has been investigated by several authors$^{3)}$. On
the other hand, here we investigate the deviation at higher
energies and the possible connection to FGT.

We can illustrate these concepts with a simple scheme of an FGT. Assume that
at energies $E \geq \Lambda_{\rm FGT}$ we have no Higgses, but a new gauge
boson $B_\mu$, with a mass $M_B~(> \Lambda_{\rm FGT})$, i.e. the symmetry
group of the extended gauge theory is $G_{\rm SM} \times G_{\rm FGT}$. The
SM--part $G_{\rm SM}$ is $SU(3)_C \times SU(2)_L \times U(1)_Y$ without
Higgses, and hence with (as yet) massless gauge bosons and fermions at these
energies. The FGT--part of Lagrangian  in the fermionic
sector is written schematically as
$$
{\cal{L}}^{\rm FGT}_{G-f} = -g \Psi \gamma^\mu B_\mu \Psi ~~~({\rm for}~~E >
\Lambda_{\rm FGT})~,
\eqno(9)
$$
where $\Psi$ is the column of all fermions and $B_\mu=B_\mu^j T_j$, $T_j$'s
being the generator  matrices of the new symmetry group $G_{\rm FGT}$.
Furthermore, we assume that the $T_j$'s corresponding to the electrically
neutral $B_\mu^j$'s do not mix flavors (i.e. no FCNC at tree level) and
are proportional to identity matrices in the flavor space. We will show in
the following lines that the FGT Lagrangian (9) can imply the creation of
SM--Higgs particles through DSB (i.e. condensation of fermion pairs),
as well as the creation of Yukawa couplings with a flavor democracy.

The effective current--current interaction, corresponding to the exchange
of neutral gauge boson $B$ at low energies $E$
($\Lambda_{\rm FGT} \leq E < M_B$), is
$$
{\cal{L}}^{\rm FGT}_{4f} \approx -{g^2 \over 2 M_B^2} \sum_{i,j}
(\bar f_i \gamma^\mu f_i)(\bar f_j \gamma_\mu f_j)~~~({\rm for}~~
\Lambda_{\rm FGT} \leq E < M_B)~.
\eqno(10)
$$
Since we are interested in the possibility of Yukawa interactions of SM
being contained in (10), and since such interactions connect left--handed
to right--handed fermions, we have to deal only with the left--to--right
(and right--to--left) part of (10). By applying Fierz transformation$^{4)}$
to this part, we end up with the 4--fermion interactions without
$\gamma_\mu$'s
$$
{\cal{L}}^{\rm FGT}_{4f} \approx {2 g^2 \over M_B^2} \sum_{i,j}
(\bar f_{iL} f_{jR})(\bar f_{jR} f_{iL})~~~({\rm for}~~\Lambda_{\rm FGT}
\leq E < M_B)~.
\eqno(11)
$$
These interactions can be rewritten in a mathematically equivalent (Yukawa)
form with auxiliary (i.e. as yet non--dynamical) scalar fields. One
possibility is to introduce only one isodoublet auxiliary scalar $H$ with
(as yet arbitrary) bare mass $M_H$, by employing a familiar mathematical
trick$^{5)}$
$$
\eqalign{
{\cal L} \approx - M_H {\sqrt{2} g \over M_B} \sum_{i,j=1}^{3}
\Biggl( &\left[ (\bar\psi^q_{iL} \tilde H) u^q_{jR} + (\bar\psi^l_{iL}
\tilde H) u^l_{jR} +h.c. \right]  \cr
+ &\left[ (\bar\psi^q_{iL} H) d^q_{jR} + (\bar\psi^l_{iL} H) d^l_{jR} + h.c.
\right] \Biggr) \cr - &M_H^2 (H^\dagger H)~~, \cr}
\eqno(12a)
$$
$$
\eqalign{
{\rm where}~~~ H &= \pmatrix{H^\dagger \cr H^0 \cr}~,~~~\tilde H = i \tau_2
H^*~~, \cr
\psi^q_i &= \pmatrix{u^q_i \cr d^q_i \cr}~,~~~\psi^l_i =
\pmatrix{u^l_i \cr d^l_i \cr}~~, \cr
u^q_i = \pmatrix{u \cr c \cr t \cr}~,
u^l_i &= \pmatrix{\nu_e \cr \nu_\mu \cr \nu_\tau \cr}~,d^q_i =
\pmatrix{d \cr s \cr b \cr}~,
d^l_i = \pmatrix{e^- \cr \mu^- \cr \tau^- \cr}~~. \cr}
\eqno(12b)
$$
Another possibility is to  introduce two auxiliary scalar isodoublets
$H^{(U)},~H^{(D)}$, with (as yet) arbitrary bare masses
$M_H^{(U)},~ M_H^{(D)}$,  and express (11) in the 2-Higgs `Yukawa' form
$$
\eqalign{
{\cal{L}} \approx &- M_H^{(U)} {\sqrt{2} g \over M_B} \sum_{i,j=1}^{3}\left[
(\bar\psi^q_{iL} \tilde H^{(U)}) u^q_{jR} + (\bar\psi^l_{iL} \tilde H^{(U)})
u^l_{jR} +h.c. \right]\cr
 &- M_H^{(D)} {\sqrt{2} g \over M_B} \sum_{i,j=1}^{3} \left[ (\bar\psi^q_{iL}
H^{(D)}) d^q_{jR} + (\bar\psi^l_{iL} H^{(D)}) d^l_{jR} + h.c. \right]
\cr &- {M_H^{(U)}}^2 ({H^{(U)}}^\dagger
H^{(U)}) - {M_H^{(D)}}^2 ({H^{(D)}}^\dagger H^{(D)})~~. \cr}
\eqno(13)
$$
As derived, the Yukawa couplings (12a) and (13) are valid in the FGT--low
energy region $(\Lambda_{\rm FGT} \leq E < M_B)$, having non--dynamical
scalar fields and being mathematically equivalent to (11). However, with
decreasing the energy, the scalars in (12) and (13) can be shown to obtain
vacuum expectation values (VEV's) and kinetic terms through quantum effects.
The neutral components of the physical Higgs doublets created through DSB
in this way are then condensates$^{6)}$ of fermion pairs
$\langle \bar f f \rangle$.
(Note that this scheme of DSB is just one of many possible scenarios for an
FGT--SM transition. However, our results will not depend on any
such specific scenario, but rather on the assumption of FD near the
transition energies, as expressed in (12) and (13).)
This would then lead us to the minimal
SM (with one Higgs) in the case (12) and to SM with two Higgses (type II)
in the case (13). Hence, although (12) and (13) are mathematically equivalent,
they lead to two physically different theories$^{2)}$.

Note that (12) implies that the minimal SM, if it is to be replaced by FGT
at high energies, should show up a trend of the Yukawa coupling matrix
(or equivalently: of the mass matrix in the flavor basis) toward a
complete flavor democracy for all fermions, with a common overall factor,
as the energy is increased within the minimal SM toward transition region
denoted as $E_0 (\equiv E_{\rm trans.})$
$$
M^{(U)}~~{\rm and}~~M^{(D)} \rightarrow {1 \over 3} m_t^0
\pmatrix{N_{FD}^q & 0 \cr 0 & N_{FD}^l \cr}~~~{\rm as}~~ E \uparrow E_0~~,
\eqno(14a)
$$
where $m_t^0=m_t(\mu=E_0)$ and
$N_{FD}$ is the $3 \times 3$ flavor--democratic matrix
$$
N_{FD}^f = \pmatrix{1 & 1 & 1 \cr 1 & 1 & 1 \cr 1 & 1 & 1 \cr}~~,
\eqno(14b)
$$
with the superscript $f=q$ for the quark sector and $f=l$ for the leptonic
sector.
On the other hand, if SM with two Higgses (type II) is to experience such
a transition, then (13) implies {\bf separate} trends toward FD for the
up--type and down--type fermions
$$
M^{(U)}~(M^{(D)}) \rightarrow {1 \over 3}~ m_t^0~(m_b^0)~
\pmatrix{N_{FD}^q & 0 \cr 0 & N_{FD}^l \cr}~~~ {\rm as}~~ E \uparrow E_0~~,
\eqno(15)
$$
where $m_t^0$ and $m_b^0$ can in general be different.
Note that $N_{FD}$, when written in the diagonal form in the mass
basis,
acquires the form
$$
N_{FD}^{\rm mass~basis} = 3 \pmatrix{0 & 0 & 0 \cr 0 & 0 & 0 \cr
0 & 0 & 1 \cr}~~.
\eqno(16)
$$
Hence, FD (and FGT) implies in the mass basis as $E$ increases to $E_0$,
$$
\eqalignno{
{m_u \over m_t},~{m_c \over m_t},~{m_{\nu_e} \over m_{\nu_\tau}},~
{m_{\nu_\mu} \over m_{\nu_\tau}} &\rightarrow 0~~, &(17a) \cr
{m_d \over m_b},~{m_s \over m_b},~~{m_e \over m_\tau},~~
{m_\mu \over m_\tau} &\rightarrow 0~~, &(17b) \cr
{m_{\nu_\tau} \over m_t},~~{m_\tau \over m_b} &\rightarrow 1~~, &(17c) \cr}
$$
and in the case of the minimal SM {\bf in addition}
$$
{m_b \over m_t},~{m_\tau \over m_{\nu_\tau}} \rightarrow 1~~.
\eqno(18)
$$
In Ref. 2, we have shown that the minimal SM does not show the required trend
toward FD, but
that SM with two Higgs doublets (type II) does.  For more details, see Ref. 2.

\medskip
\leftline{\bf 3. HADRONIC INVARIANT MASS DISTRIBUTION ON SEMILEPTONIC}
\centerline{\bf $B$--DECAY}
\medskip

Semileptonic decays of $B$-mesons continue to attract great
experimental and theoretical interest, because of
the light they may throw  both on heavy quark  decay mechanisms and on KM
matrix elements. Theoretical approaches have been made both from the inclusive
point of view, as in the independent-quark decay model of Altarelli
\etal$^{7)}$, and also by calculating and summing exclusive channels as
with Wirbel \etal$^{8)}$ and others. Here we adopt the
inclusive approach.

In the independent-quark decay model (or ``spectator model'') of Ref.~7, the
initial $B$-meson is represented by a pair of quarks $b+\bar q \
(\bar q = \bar u$ or $\bar d$) with a distribution of Fermi momentum
${\bf p} = {\bf p} (\bar q) =-{\bf p} (b)$ in the $B$ rest-frame. Then $b$
decays as a free quark via $b
\to q' \ell\nu$ (with $q'=c$ or $u$); $\bar q $ remains an independent
spectator. The spectator mass is assigned a value of order $m_{\rm sp} \simeq
0.1$~GeV (the precise value is not critical). Kinematical constraints then
require the effective mass of the decaying $b$-quark to depend on the Fermi
momentum {\bf p}:
$$m^2_b =m^2_B +m^2_{\rm sp} -2m_B (m^2_{\rm sp} +p^2)^{1/2} ,\eqno(19)$$
where $p = |{\bf p}|$.
 The primary role of the Fermi momentum is to reflect the initial $B(b\bar q)$
wave function, through the impulse approximation; however, in the spectator
picture, it also subsumes some effects of final state interactions. For
example, large values of $p$ would contribute toward broad distributions in the
final hadronic invariant mass $m(\bar q q')$; if final-state interactions
enhance a
relatively low-mass region, their effects could appear as an enhancement of the
effective Fermi momentum distribution at small $p$.

The lepton spectrum in $b \to c\ell\nu$ decays has the form
$${d\Gamma(b \to c\ell\nu) \over dx_\ell} ={d\Gamma^0 \over dx_\ell} \left[ 1 -
{2\alpha_s \over 3\pi} G(x_\ell, \epsilon) \right] \eqno(20)$$
where $x_\ell =2(\ell\cdot b) /m^2_b =2E_\ell/m_b$ in the $b$ rest-frame,
$\Gamma^0$
is the standard lowest-order electroweak calculation, $\alpha_s$ is the QCD
coupling constant and $\epsilon =m_c/m_b$. The factor in square brackets is the
lowest-order QCD correction from virtual and real gluon emission and
$G(x_\ell,\epsilon)$ is a complicated function defined in Ref.~7,
including a Sudakov resummation of double logarithms to ensure a finite result
at the end-point $x_\ell (\max) \equiv x_M =1-\epsilon^2$.
$G$ depends little on $x_\ell$ except
near $x_M$; for other
values of $x_\ell, \ G$ is close to its spectrum-averaged value $g(\epsilon)$,
defined by the integrated width
$$\Gamma(b \to c\ell\nu) =\Gamma^0 (b\to c\ell\nu) \left[ 1 -{2\alpha_s\over
3\pi} g(\epsilon) \right] \,. \eqno(21)$$
There is a useful empirical approximation$^{9)}$ accurate within
0.2\%,
$$g(\epsilon) \simeq (\pi^2 - 31/4) (1-\epsilon)^2 +3/2 \,. \eqno(22)$$
For completeness we note that
$$\eqalignno{\Gamma^0(b \to c\ell\nu) &={G^2_F m^5_b \over 192\, \pi^3}
|V_{cb}|^2
(1-8\epsilon^2 +8\epsilon^6 -\epsilon^8 -24\, \epsilon^4 \ln\epsilon) \,,
&(23) \cr \noalign{\vskip.1in}
{d\Gamma^0 \over dx_\ell} (b\to c\ell\nu) &= {G^2_F m^5_b \over 96\, \pi^3}
|V_{cb}|^2 x^2_\ell {(1 \! - \! \epsilon^2 \! - \! x_\ell) ^2 \over (1 \! - \!
x_\ell)^3}
\left[ (1 \! - \! x_\ell) (3 \! - \! 2x_\ell) + \epsilon^2 (3 \! - \! x_\ell)
\right] , \ &(24)\cr }$$
where $V$ is the CKM matrix and lepton mass is ignored. For each value of the
Fermi momentum ${\bf  p}$, we calculate $d\Gamma/dE_\ell$ in the $b$
rest-frame (isotropic angular distribution) and boost to the $B$ rest-frame
folded with Fermi motion ${\bf p}$. For comparison with data at the
$\Upsilon(4S)$ resonance, where $B$ is produced not precisely at rest, a
further boost to the lab frame is made with appropriate angular integrations.

The CKM matrix element $V_{ub}$ characterizing $b\to u$ quark transitions plays
an important role in the description of CP violation within the three-family
Standard Model, but is still not accurately known. The most direct way to
determine this parameter is through the study of $B$ meson semileptonic decays;
recent results from the CLEO$^{10)}$   
and ARGUS$^{11)}$ 
data on the
end-point region of the lepton spectrum have established that $V_{ub}$ is
indeed non-zero and have given an approximate value for its modulus.
The central problem in the extraction of $V_{ub}$ is
the separation of $b\to u$ events from the dominant $b\to c$ events. In
semi-leptonic $B$-meson decays, the usual approach  
 is to study the upper
end of the charged lepton spectrum, since the end-point region
$$
E_\ell > (m_B^2 - m_D^2 + m_\ell^2)/(2m_B)
\eqno(25)
$$
in the CM frame is inaccessible to $b\to c$ transitions and therefore selects
purely $b\to u$. However, only about 20\% of $b\to u$ transitions actually
lie in the region of Eq.~(25); it is therefore not a very efficient way to
select them. In situations of physical interest, the situation is even
somewhat worse.  For example, in $\Upsilon(4S) \to  B \bar B$ decay, each
$B$ meson has a small velocity in the $\Upsilon$ rest frame; the magnitude
$\beta$ of this velocity is known, but its direction is not. In this frame,
which is the laboratory frame when $\Upsilon$ is produced at a symmetric
$e^+e^-$ collider, the $b\to u$ selection region based on Eq.~(25) becomes
$$
E_\ell > \gamma(1+\beta)(m_B^2 - m_D^2)/(2m_B) ,
\eqno(26)
$$
for the cases $\ell=e$ or $\mu$.     Here $\gamma = (1-\beta^2)^{-1/2} =
m_\Upsilon/(2m_B)$  and we neglect the lepton mass.  (At an asymmetric
collider, where $e^+$ and $e^-$ beams have different energies, it will be
necessary to boost lepton momenta from the laboratory frame to the
$\Upsilon$ rest frame before applying this cut.)  Equation~(26) accepts an even
smaller percentage of $b\to u$ decays than Eq.~(25), about 10\% in fact.

The essential physical idea behind Eqs.~(25)-(26) is that $b\to c$ transitions
leave at least one charm quark in the final state; hence for a general
semileptonic decay  $B \to  \ell + \nu + X $  the invariant mass $m_X$
of the final hadrons exceeds $m_D$ and this implies a kinematic bound on
$E_\ell$.
In Ref. 12, we give this old idea a new twist. We first observe that
there is no unique connection
between $m_X$ and $E_\ell$, due to the presence of the neutrino, so the bound
on $E_\ell$ is not an efficient way of exploiting the bound on $m_X$.
We then observe a more efficient way to exploit the
latter bound is to measure $m_X$ itself and to select $b\to u$ transitions
by requiring
$$
m_X < m_D
\eqno(27)
$$
instead of Eqs.~(25)-(26). This condition is of course frame-independent.

The final hadronic invariant mass distribution depends both on the $c$-quark
energy distribution $d\Gamma(b \to c\ell\nu)/dE_c$ and on the Fermi momentum
distribution $\phi(p)$ which is normalized to $\int^\infty_0 dp \, \phi(p) =1$
and is sometimes approximated by a gaussian form
$$\phi(p) =4p^2(p^3_F\sqrt\pi)^{-1} {\rm exp}(-p^2/p^2_F)~, \eqno(28)$$
where  $p_F$ is a characteristic parameter describing Fermi motion
inside $B$.  The lowest-order contribution to the $c$-quark energy
distribution is given by
$${d\Gamma^0(b \to c\ell\nu) \over dx_c} = {G^2_F m^5_b \over 96\, \pi^3}
|V_{cb}|^2 (x^2_c -4\epsilon^2)^{1/2} \left[ 3x_c (3-2x_c) + \epsilon^2 (3x_c -
4) \right] , \eqno(29)$$
where $x_c = 2(c\cdot b)/m^2_b=2E_c/m_b$ in the $b$ rest-frame, with
kinematical
range $2\epsilon \le x_c \le 1+\epsilon^2$.
When QCD radiative corrections are included, the real and virtual gluon
contributions must be subject to resolution smearing so that their singular
parts will cancel; this we approximate by absorbing real soft gluons into the
effective final $c$-quark and correcting $d\Gamma^0/dx_c$ by the factor
$g(\epsilon)$:
$${d\Gamma(b \to c\ell\nu) \over dx_c} \simeq {d\Gamma^0(b\to c\ell\nu) \over
dx_c } \left[ 1 -{2\alpha_s \over 3\pi}  g(\epsilon) \right] .  \eqno(30)$$
For each value of the Fermi momentum $p$ we calculate $d\Gamma/dE_c$ in the $b$
rest-frame (isotropic angular distribution here) and fold it with the spectator
energy-momentum vector to form the distribution $d\Gamma/dm_X$ with respect to
the invariant mass $m_X$ of the final charmed hadronic system,
$$m^2_X =(E_c + E_{\rm sp})^2 - ({\bf p}_c + {\bf p}_{\rm sp})^2. \eqno(31)$$
The spectator energy and momentum in the $b$ rest-frame are
$$\eqalign{E_{\rm sp} &= \left[ (p^2 +m^2_b)^{1/2} (p^2 +m_{\rm sp}^2)^{1/2}
+p^2 \right] /m_b \,, \cr
{\bf p}_{\rm sp} &= \left[ (p^2 +m^2_b)^{1/2} + (p^2 + m^2_{\rm sp})^{1/2}
\right] {\bf p} /m_b
\,, \cr} \eqno(32)$$
and $m_b$ is everywhere defined by Eq.~(19). The maximum and minimum
values of $m^2_X$ for given $p$ are
$$\eqalign{m^2_X(\max) &= m^2_c +m^2_{\rm sp} + m_b(E_{\rm sp} +p_{\rm sp})
+m^2_c(E_{\rm sp} -p_{\rm sp})/m_b \,, \crr
m^2_X(\min) &=
\cases{(m_c +m_{\rm sp})^2 \,, \qquad\qquad\qquad
 {\rm if} \ \  (m^2_b -m^2_c) E_{\rm sp}
\ge (m^2_b +m^2_c) p_{\rm sp}, &\cr \noalign{\vskip.1in}
m^2_c +m^2_{\rm sp} +m_b(E_{\rm sp} -p_{\rm sp}) + m^2_c(E_{\rm sp} +p_{\rm
sp})/m_b, \quad \ {\rm  otherwise}. \cr}
\cr}\eqno(33)$$
These relations show explicitly that small $p$ results in $m_X$ values close to
$(m_c +m_{\rm sp})$. The upper limit on $p$ for the decay to be possible (from
Eq.~(19) with $m_b >m_c$) is
$$p < \lambda^{1/2} (m^2_B, m^2_c, m^2_{\rm sp})/(2m_B) \,, \eqno(34)$$
where $\lambda(a,b,c) =a^2 +b^2 +c^2 -2ab -2bc -2ac$.

For $b \to u$ transitions the effect of individual resonances in $X$ quickly
disappears above the $\pi$ and $\rho$ region, and multiparticle jet-like
continuum final states should give the dominant contributions$^{13)}$;
this makes
it reliable to calculate the $m_X$ distribution using the modified spectator
decay model.     Figure~1 gives the hadronic
invariant mass distribution from $B\to \ell \nu X$ semileptonic decays,
showing that more than
90\% of $b\to u$ decays lie within the region selected by Eq.~(27).  In this
illustration we use $m_B =5.273$~GeV, $E_B
=m_\Upsilon/2 =5.29$~GeV, $m_c=1.6$~GeV, $m_u = 0.1$~GeV,
$p_F=0.3$~GeV,
including QCD corrections up to order $\alpha\alpha_s$ according to Ref.~14.
The figure is normalized for simplicity to the case $|V_{ub}/V_{cb}|=1$.

In order to exploit Eq.~(27) instead, it is desirable to  isolate
uniquely the products of a single $B$ meson decay; to achieve this
it is generally necessary to reconstruct both $B$ decays in a given
event.  One of these decays can be semileptonic, since kinematic
constraints can often determine the missing neutrino four-momentum well
enough to reconstruct a peak at zero in the invariant square of this
four-momentum. Double-semileptonic decay events will not generally
reconstruct uniquely, however. To study semileptonic channels, we are
therefore concerned with those events (about 30\% of the total) where
one $B$ decays hadronically, one semileptonically with $\ell=e$ or $\mu$.
Of order 1\% of these have  $b\to u \ell\nu$  semileptonic transitions (because
of
the very small ratio $|V_{ub}/V_{cb}| \approx 0.1$).
About 10\% of the latter satisfy the criterion  $E_\ell > 2.5$~GeV  of
Eq.~(26).
With present data, it appears to be possible to reconstruct a few
percent of such events, but perhaps only about one percent without
ambiguity. The arithmetic of collecting reconstructed $b\to u\ell\nu$ events
with the criterion of Eq.~(26) is then approximately as follows:
$$
\eqalign{
10^6 \, B \bar B\ {\rm  events}
 &\to  3 \times 10^5 {\rm \  events \ with \ one \ } (b\to q\ell\nu)\cr
 &\to  3 \times 10^3 \ {\rm events \ with \ one} \  (b\to u\ell\nu)\cr
 &\to  3 \times 10^2 \ {\rm events \ with \ } (b\to u\ell\nu, \
E_\ell>2.5~{\rm GeV}) \cr
 &\to  3 \ {\rm reconstructed} \ (b\to u\ell\nu,\  E_\ell>2.5~{\rm GeV}). \cr}
\eqno(35)
$$
The numbers in Eq.~(35) are at first sight discouraging to hopes of an
extensive study of  $b\to u\ell\nu$  transitions.

\medskip
\noindent {\it Figure 1.}  Hadronic invariant mass distribution in $B \to \ell
\nu X$
semileptonic decays.
\vfill\eject

In contrast, using the criterion of Eq.~(27) instead offers an order of
magnitude more events:
$$
\eqalign{
10^6 \, B \bar B\ {\rm  events}
 &\to  3 \times 10^3 {\rm \  events \ with \ } (b\to u\ell\nu, \ m_X < m_D)
\cr
 &\to  30 \ {\rm reconstructed} \ (b\to u\ell\nu,\  m_X < m_D).\cr}
\eqno(36)
$$
This is more encouraging.
Also, the introduction of microvertex detectors
(ARGUS has recently added one) can be expected to improve the success
rate for reconstructing $B\bar B$ events and to reduce the ambiguous
cases. For symmetrical colliders, microvertex detectors will assist in
identifying charm decay vertices and the charged tracks associated
with them.  For projected future asymmetrical colliders, where the two
beams have different energies and the produced $\Upsilon(4S)$ is not at
rest in the laboratory, the bottom decay vertices too will often be
identifiable, with still greater advantage to the analysis. These
future ``$b$-factories'' will also deliver much larger numbers of events;
samples of order $10^8$ or more $B\bar B$ events are foreseen, in which case
the numbers in Eq.~(36) can be scaled up by at least two orders of magnitude.

We note that there is a question of bias.  Some classes of final
states (e.g. those with low multiplicity, few neutrals) may be more
susceptible to a full and unambiguous reconstruction. Hence an
analysis that requires this reconstruction may be biassed. However
the use of topological information from microvertex detectors
should tend to reduce the bias, since vertex resolvability depends
largely on the proper time of the decay and its orientation relative
to the initial momentum (that are independent of the decay mode).
Also such a bias can be allowed for in the analysis, via suitable
modeling.
Finally there may be a background from
continuum events that accidentally fake the $\Upsilon$ events of interest.
This can be measured directly at energies close to the resonance.

\medskip
\leftline{\bf 4. TOPICS ON $K$ AND $B$ PHYSICS}
\medskip
\leftline{\bf 4.1 Role of vector mesons in rare kaon decays$^{15)}$}
\medskip

\def\kpgg{K_{L} \rightarrow \pi^{0} \gamma \gamma}
\def\kpp{K \rightarrow \pi \pi}
\def\deli{\Delta I = 1/2}
\def\kpll{K \rightarrow \pi l^+ l^-}
\def\kgll{K_{L} \rightarrow \gamma l^+ l^-}
\def\kgee{K_{L} \rightarrow \gamma e^+ e^-}

\def\kpee{K_{L} \rightarrow \pi^{0} e^+ e^-}
\def\gg{\gamma \gamma}
\def\kppee{K^{+} \rightarrow \pi^{+} e^+ e^-}

\def\kppmu{K^{+} \rightarrow \pi^{+} \mu^{+} \mu^{-}}
\def\ksee{K_{S} \rightarrow \pi^{0} e^+ e^-}
\def\ksmu{K_{S} \rightarrow \pi^{0} \mu^{+} \mu^{-}}
\def\kgg{K_{L} \rightarrow \gamma \gamma}

The role of vector mesons in $\kpgg$ and $\kpee$ has
been a controversial subject$^{16,17)}$.  Chiral perturbation
theory$^{16)}$ and the
pion rescattering model$^{18)}$ predict that the pion loop gives a
dominant
contribution to $\kpgg$ with branching ratio around $7 \times
10^{-7}$, and the two photon spectrum has a peak near
$m_{\gg} \simeq 300$ MeV in both models.  Also, the low energy photon
pair is almost negligible.  On the other hand, there can be a large
enhancement in the low $m_{\gg}$ region$^{19,20)}$,
if vector mesons come into
play in this decay mode and one uses the naive nonet symmetry in $K_{2}
\rightarrow P$ (with $P = \pi^{0}, \eta_{8}, \eta_{0}$).
In this case, the branching ratio is $(1 \sim 3) \times 10^{-6}$.
The recent measurement$^{21)}$ of $\kpgg$
from CERN is rather
puzzling. The two photon spectrum seems consistent with the
predictions of the chiral perturbation theory  and the pion
rescattering model.  However, the branching ratio for $m_{\gg} \ge 280$
MeV is larger
than those predictions by a factor of 3 $\sim$ 4.
In this rather confusing situation, it would be nice to have a
systematic calculation for $\kpgg$ as well as for other processes
where
vector meson contributions can be potentially important.

In  Ref. 15, the author gives self-consistent calculations for
$\kpp$, $\kpll$, $\kgg$, $\kgll$ and $\kpgg$ in the hidden symmetry
scheme$^{22)}$ in
conjunction with the nonleptonic Hamiltonian used by  J.J.
Sakurai, J.A. Cronin and other groups$^{23)}$.
In his approach, the usual $O(p^4)$ and $O(p^{6})$ terms
arise from vector meson
exchange between the chiral mesons, and from the Wess--Zumino anomaly
term including vector mesons.  From the work of Ref.~20, it is
known that the $O(p^{6})$ terms through $\rho^{0}$ and $\omega$
exchange in $\gamma \gamma \rightarrow \pi^{0} \pi^{0}$
unitarize the chiral loop amplitude in an effective way up to
$m_{\gamma \gamma} \sim 1$ GeV, and controls the high energy behavior
of the chiral amplitudes.
Once we make this assumption, things get much simplified and we
anticipate we do not lose any important physics information.
The major advantage of our model lies in the number of unknown
parameters to be
determined by the experimental data.  In fact, we do not have
any unknown parameters at the level of strong and electromagnetic
interactions
other than the meson masses and their decay
constants, {\it once} we adopt the notion of vector meson
dominance in the normal sector. We first consider the structure of the
left--handed
currents.
We find that there exist anomalous left--handed currents arising from the
Wess--Zumino anomaly and the intrinsic parity violating interactions
involving vector mesons, and they give important contributions to $\kgll$ and
$\kpgg$ through generating  {\it weak} $V \pi \gamma$ vertices.
The nonleptonic weak decays of kaons is described by the effective
weak Lagrangian of current--current interactions.  We make connections
with the calculations in terms of quark fields with QCD
corrections$^{24,25)}$,
and find that there is an addtional term in the  effective weak
Lagrangian, which is proportional to Tr ($j_{\mu L}$) where `Tr' is taken
over the $U(3)_{f}$ or $SU(3)_{f}$ indices.  This additional term
cannot be thrown away as usually done
in the case of $SU(3)_{L} \times SU(3)_{R}$,
since the anomalous
left--handed currents we consider here have nonvanishing trace even
in that case.
The coefficient of this new term measures the contributions of
penguin operators of current--current types to the nonleptonic kaon
decays.
At this stage, we will have four parameters, $C_8^{(1/2)},
C_{27}^{(1/2)},
C_{27}^{(3/2)} $ and $\delta_p$ with one constraint
$C_{27}^{(3/2)} = 5 C_{27}^{(1/2)}$.
$C$'s characterize the strengths of
the $(8_{L},1_{R})_{\deli}$, $(27_{L},1_{R})_{\deli}$ and
$(27_{L},1_{R})_{\Delta I = 3/2}$ pieces of the weak Hamiltonian,
respectively. Deviation of $\delta_p$ from 1 measures the
contributions of penguin operators.
We fix $C$'s from $\kpp$.  $\delta_p$ contributes to $\kpgg$ in
the $SU(3)_{L} \times SU(3)_{R}$ case, and both to $\kpgg$ and to $\kgll$
in the $U(3)_{L} \times U(3)_{R}$ case.

Next we assume nonet symmetry in the vector meson sector.
Before studying the effects of $\delta_p$ on $\kgll$,
we analyze $\kpll$  using the suitable value of
$\delta_p$.  For the decay mode $\kpll$,
we need two more operators,
$Q_{7 V} \equiv (\bar s d)_{V-A} (\bar l l)_{V}$ and $Q_{7 A} \equiv (\bar s
d)_{V-A} (\bar l l)_{A}$,
arising from the electromagnetic penguin diagram$^{26)}$,
the $Z^0$ penguin diagram
and the box diagram with two internal $W$'s$^{27)}$.  The real part of
the Wilson coefficients
of these operators cannot be calculated reliably. Therefore, we introduce
another parameter $C_7$ as the coefficient of the new operator $Q_{7 V}$,
ignoring the operator $Q_{7 A}$.
We fix $C_7$ from the best fit to the decay mode $\kppee$.  There is a
twofold ambiguity in $C_7$, and we can predict for $\kppmu$, $\ksee$
and $\ksmu$.  This model predicts the decay rate
for $\ksee$ process to be comparable to the decay rate for $\kppee$.
The $\ksee$ process contributes to the indirectly  CP--violating
$\kpee$
process through the mixing between $K_{L}$ and $K_{S}$.  We predict
that
the branching ratio of the indirectly  CP--violating $\kpee$ process
is about 1.4 or 2.7 $\times 10^{-10}$,  which is substantially
larger than other previous calculations.
Another process in which vector mesons are important is $\kgee$.
This has been recently remeasured and the form factor shows a clear
deviation from the $\rho$ form factor$^{28,29)}$.  Then we make an
analysis of $\kgg$ and $\kgll$.
In $\kgll$, we have both {\it weak} $VV$ and {\it weak} $V \pi \gamma$
vertices, the latter of which was not considered in the earlier
analysis.  Next we introduce one more parameter $\delta_n$
characterizing the possible deviation of $a(K_{2} \eta_{0})$ from its
naive value obtained from the effective weak Lagrangian.  This takes
care of the fact that the $U(1)_A$ symmetry is broken through the QCD
axial anomaly.
The recent data on $\kgee$, when combined with the branching ratio of
$\kgg$, provide us with important information
on $\zeta \equiv a(K_{2} \eta )/ a(K_{2} \pi^{0})$,
$\zeta^{'} \equiv  a(K_{2} \eta^{'} )/ a(K_{2} \pi^{0})$ and
$\delta_p$, or equivalently, on $\delta_n$ and $\delta_p$.
We will have two solutions for $\delta_n$, each of which corresponds to
$\pi$-- and $\eta$-- dominance in $\kgg$,
respectively.
Then, $\delta_p$
is constrained to some region for each $\delta_n$.
In $\kppee$,
we could determine $C_{7} = -0.01_{-0.04}^{+0.03}$ or $- 0.61_{-0.04}^{+0.03}$
from the best fit to
$B(\kppee)$.  This parameter measures the short distance contribution
of the electromagnetic penguin diagram to the process $\kpll$.
The twofold ambiguity can be lifted by the
spectrum measurement of $\kppee$ at low $m_{ee}$.
For either value of $C_{7}$,
we are led to rather large branching ratios for the
indirect CP--violating $\kpee$ at the level of a few parts in $10^{-10}$.

\medskip
\leftline{\bf 4.2 $b \to s \gamma$ in MSSM and flipped $SU(5)$ SUGRA$^{30)}$}
\medskip

\def\ie{\hbox{\it i.e.}}

\def\coeff#1#2{{\textstyle{#1\over #2}}}

\catcode`\@=11 

\def\lsim{\mathrel{\mathpalette\@versim<}}
\def\gsim{\mathrel{\mathpalette\@versim>}}
\def\@versim#1#2{\vcenter{\offinterlineskip
    \ialign{$\m@th#1\hfil##\hfil$\crcr#2\crcr\sim\crcr } }}
\def\boxit#1{\vbox{\hrule\hbox{\vrule\kern3pt
      \vbox{\kern3pt#1\kern3pt}\kern3pt\vrule}\hrule}}

\def\etal{{\it et. al.}}

\def\t1{{\tilde 1}}

\def\JL{J. L. Lopez}
\def\DVN{D. V. Nanopoulos}
\def\AZ{A. Zichichi}
\def\HP{H. Pois}

\def\GeV{\,{\rm GeV}}

\def\bsg{b\to s\gamma}
\def\brbsg{{\rm BR}(\bsg)}

The purpose of this subsection is to study $\brbsg$ in two supergravity models:
(i) the minimal $SU(5)$ supergravity model$^{31)}$, and (ii) the no-scale
flipped $SU(5)$ supergravity model$^{32)}$. We find that the {\it new} CLEO
bound$^{33)}$, i.e. BR($b \to s \gamma) < 5.4\times10^{-4}$ at 95\% CL,
 does not yet constrain the minimal $SU(5)$ model, while some as-yet-mild
constraints are imposed on the flipped $SU(5)$ model, where a new phenomenon
can drastically suppress the $\bsg$ amplitude. We present the results for
$\brbsg$ in these two models and show that improved sensitivity could probe
them in ways not possible at present collider experiments.

We use the following expression$^{34)}$ for the branching ratio $\bsg$
$$
{\brbsg\over{\rm BR}(b\to ce\bar\nu)}={6\alpha\over\pi}
{\left[\eta^{16/23}A_\gamma
+\coeff{8}{3}(\eta^{14/23}-\eta^{16/23})A_g+C\right]^2\over
I(m_c/m_b)\left[1-\coeff{2}{3\pi}\alpha_s(m_b)f(m_c/m_b)\right]},
\eqno(37)
$$
where $\eta=\alpha_s(M_Z)/\alpha_s(m_b)$, $I$ is the phase-space factor
$I(x)=1-8x^2+8x^6-x^8-24x^4\ln x$, and $f(m_c/m_b)=2.41$ the QCD
correction factor for the semileptonic decay. The $A_\gamma,A_g$ are the
coefficients of the effective $bs\gamma$ and $bsg$ penguin operators
evaluated at the scale $M_Z$. Their simplified expressions are given in
the Appendix of Ref. 34, where the gluino and neutralino contributions have
been justifiably neglected$^{35)}$ and the squarks are considered
degenerate in mass, except for the $\tilde t_{1,2}$ which are significantly
split by $m_t$. This is a fairly good approximation to the actual result
obtained in the two supergravity models we consider here, since $m_{\tilde
q}>200\GeV$ in these  models. To include at least partially the QED corrections
to the $\bsg$ operator, in Eq. (37) we take $\alpha=\alpha(m_b)=1/131.2$.
We use the 3-loop expressions for $\alpha_s$ and choose $\Lambda_{QCD}$ to
obtain $\alpha_s(M_Z)$ consistent with the recent measurements at LEP.
For the numerical evaluation of Eq. (37) we take $\alpha_s(M_Z)=0.118$,
${\rm BR}(b\to ce\bar\nu)=10.7\%$, $m_b=4.8\GeV$ and $m_c/m_b=0.3$.

The subject of the QCD corrections to $\brbsg$, \ie, the origin of the $\eta$
factors and the $C$-coefficient in Eq. (37), has received a great deal of
attention over the years in the SM. We use the leading-order QCD
corrections to the $\bsg$ amplitude when evaluated at the $\mu=m_b$ scale,
\ie, $C=\sum_{i=1}^8 b_i\eta^{d_i}=-0.1766$ for $\eta=0.548$, with the
$b_i,d_i$ coefficients given in Ref. 34.  The result follows from the
renormalization-group scaling from the scale $M_Z$ down to $\mu=m_b$ of the
effective $\bsg$ operators at $M_Z$, which include the usual electromagnetic
penguin operator plus some four-quark operators. Scaling introduces operator
mixing effects (as exemplified by the appearance of the gluonic penguin
operator in Eq. (37)) and the scaled coefficients at $\mu=m_b$ are given by
linear combinations of the coefficients at scale $M_Z$. It has been pointed
out$^{36)}$ that the low-energy mass scale $\mu$ affects the leading-order
results
significantly. The combined effect on
$\brbsg$ of the uncertainties in $\mu$ and $\Lambda_{QCD}$ has been
estimated$^{34)}$
to be $\lsim25\%$. Recently a partial next-to-leading order
calculation of the QCD effects has appeared$^{37)}$, which gives somewhat
smaller enhancement factors than the complete leading-order result. A full
next-to-leading order QCD calculation should decrease the above mentioned
uncertainties significantly.

The two models we consider here are built within the framework of supergravity
with
universal soft-supersymmetry breaking. The renormalization-group scaling
from the unification scale down to low energies plus the requirement of
radiative electroweak symmetry breaking using the one-loop effective potential,
reduces the number of parameters needed to describe these models down to
just five: $m_t$, $\tan\beta$, and three soft-supersymmetry breaking parameters
($m_{1/2},m_0,A$). The sign of the superpotential Higgs mixing term $\mu$
remains as a
discrete variable. The models we consider belong to this general
class of models but are further constrained making them quite predictive.

The minimal $SU(5)$ supergravity model$^{31)}$ is strongly constrained by
the proton lifetime and the cosmological constraint of a not too young
universe$^{38-40)}$. A thorough exploration of the parameter
space, including two-loop gauge coupling unification  yields a restricted
region of parameter space to be subjected to further phenomenological tests.
We find
$$
2.3\,(2.6)\times10^{-4}<\brbsg_{minimal}<3.6\,(3.3)\times10^{-4},
\eqno(38)
$$
for $\mu>0\,(\mu<0)$, which are all within the new CLEO bound. We also obtain
$$
0.90\,(0.97)<\brbsg_{minimal}/\brbsg_{SM}<1.20\,(1.13),
\eqno(39)
$$
for $\mu>0\,(\mu<0)$. This implies that $\brbsg$ would need to be measured
with better than $20\%$ accuracy to start disentangling the minimal $SU(5)$
supergravity model from the SM. Moreover, for $\mu<0$ there is a band of
points which will be difficult to tell apart (requires $<1\%$ accuracy).
It is interesting to remark that an analogous set of plots versus $m_t$
instead, does not reveal any particular structure in $m_t$.
In other words, in the minimal
$SU(5)$ supergravity model, a measurement of $\brbsg$ within the range of
Eq. (39), is unlikely to shed much light on the value of $m_t$. Of course, if
the measurement falls outside this range, the model would be excluded.

The no-scale flipped $SU(5)$ supergravity model$^{32)}$ assumes that the
parameters $m_0$ and $A$ vanish, as is typically the case in no-scale
supergravity models$^{41)}$. This constraint reduces the dimension of the
parameter
space down to three. The choice of gauge
group (flipped $SU(5)$) is made to make contact with string-inspired models
which unify at the scale $M_U\sim10^{18}\GeV$. We also consider a ``strict
no-scale" scenario where in addition the universal
bilinear soft-supersymmetry breaking scalar mass parameter $B$ vanishes. This
constraint reduces the dimension of the parameter space down to two, since
$\tan\beta$ can now be computed as a function of $m_{1/2}$ and $m_t$. An
interesting consequence of this scenario is that for $\mu>0$ only
$m_t\lsim135\GeV$ is allowed, whereas for $\mu<0$ only $m_t\gsim140\GeV$ is
allowed. Moreover, for $\mu>0$ the calculated value of $\tan\beta$ can be
double-valued. In what follows we take $m_t=100,130\,(140,150,160)\GeV$ for
$\mu>0\,(\mu<0)$.
The results are as follows: (i) if $m_t\gsim140\GeV$ then $\mu<0$ and
$\brbsg\gsim3.5\times10^{-4}$; and (ii) if $m_t\lsim135\GeV$ then $\mu>0$ and
a wide range of $\brbsg$ values are possible. In both cases, with sufficiently
accurate measurements, one should be able to pin down the value of the chargino
mass (with a possible two-fold ambiguity) and therefore the whole spectrum of
the model.

\medskip
\leftline{\bf ACKNOWLEDGEMENTS}
\medskip

I would like to thank V. Barger, G. Cvetic, F. Cuypers, A.D. Martin, R.J.N.
Phillips for the previous fruitful collaborations. Many thanks should go
to our discussion section participants, especially to K. Abe, Pyungwon Ko, B.H.
Lee,
E. Ma, and G.T. Park who gave short talks to make our section most active.
My special thanks go to the organizers of the 14th Interanational Workshop
on Weak Interactions and Neutrino, and more to S.K. Kim.
This work was supported in part by the Korean Science and Engineering
Foundation, in part by Korean Ministry of Education, in part by the Center for
Theoretical Physics, Seoul National University, and in part by Yonsei
University
Faculty Research Grant.

\medskip
\leftline{\bf REFERENCES}
\medskip

\item{1.} M. Kobayashi and T. Maskawa, {\it Prog. Theo. Phys.} 49 (1973), 652.

\item{2.} G. Cvetic and C.S. Kim, DO--TH 92--08/YSTP 92--03;
DO--TH 93--08/SNUTP 93--12/YUMS 93--05 ({\it Nucl.
Phys.} B, in press).

\item{3.} Y. Nambu, {\it Proceedings of XI Warsaw symposium on High Energy
Physics } (1988); P. Kaus and S. Meshkov, {\it Phys. Rev.}  D42 (1990), 1863;
H. Fritzsch and J. Plankl, {\it Phys. Lett}  B237 (1990), 451; F. Cuypers
and C.S. Kim, {\it Phys. Lett}  B254 (1991), 462.

\item{4.} M.A. Shifman, A.I. Vainshtein and V.I. Zakharov, {\it Nucl. Phys.}
 B120 (1977), 316.

\item{5.} T. Eguchi, {\it Phys. Rev.}  D14 (1976), 2755.

\item{6.} Y. Nambu and G. Jona--Lasinio, {\it Phys. Rev.} 122 (1961), 345;
 124 (1961), 246.

\item{7.} G.~Altarelli, N.~Cabbibo, G.~Corbo, L.~Maiani, and G.~Martinelli,
{\it Nucl. Phys.} B208 (1982), 365.

\item{8.} M. Wirbel, B. Stech, and M. Bauer, {\it Z. Phys.} C29 (1985), 637;
M. Bauer, and M. Wirbel, {\it Z. Phys.} C42 (1989), 671.

\item{9.} C.S. Kim, and A.D. Martin, {\it Phys. Lett.} B225  (1989), 186.

\item{10.} CLEO collaboration: R.~Fulton \etal, {\it Phys. Rev. Lett.} 64
(1990), 16.

\item{11.} ARGUS collaboration: H.~Albrecht \etal, {\it Phys. Lett.} B234
(1990), 409;
 B241  (1990), 278.

\item{12.} V.~Barger, C.~S.~Kim, and R.~J.~N.~Phillips, {\it Phys. Lett.} B235
(1990),
187; B251 (1990), 629.

\item{13.} C.~Ramirez, J.~F.~Donoghue, and G.~Burdman, {\it Phys. Rev.} D41
(1990), 1496.

\item{14.} M.~Jezabek and J.~H.~K\" uhn, {\it Nucl. Phys.} B320 (1989), 20.

\item{15.} A short talk given by Pyungwon Ko. For more details, see
{\it Phy. Rev.} D44 (1991), 139.

\item{16.} G. Ecker, A. Pich and E. de Rafael, {\it Phys. Lett. } B189 (1987),
363;
{\it Nucl. Phys. } B303 (1988), 665.

\item{17.} L.M. Sehgal, {\it Phys. Rev. } D38 (1988), 808; T. Morozumi
and H. Iwasaki, KEK Preprint TH-206 (1988); J. Flynn and L.
Randall, UCB-PTH-88-21, LBL-26008, RAL-88-080.

\item{18.} P. Ko and J.L. Rosner, {\it Phys. Rev.}  D40 (1989), 3775.

\item{19.} L.M. Sehgal, {\it Phys. Rev.}  D41 (1990), 161.

\item{20.} P. Ko, {\it Phys. Rev.} D41 (1990), 1531.

\item{21.} G.D. Barr {\it et al.}, {\it Phys. Lett.}  B242 (1990), 523;
also, V. Papadimitriou et al., {\it Phys. Rev. Lett.} 63 (1989), 28;
V. Papadimitriou, Thesis, University of Chicago, 1990.

\item{22.} M. Bando, T. Kugo and K. Yamawaki, {\it Phys. Rep.} 164 (1988), 217,
 and references therein.

\item{23.} J.J. Sakurai, {\it Phys. Rev.}  156 (1967), 1508;
J.A. Cronin, {\it Phys. Rev.} 161 (1967), 1483;
M. Moshe and P. Singer, {\it Phys. Rev.}  D6 (1972), 1379.

\item{24.} A.I. Vainshtein, V.I. Zakharov and M.A. Shifman, {\it Sov. Phys.
JETP} 45 (1977), 670.

\item{25.} F. Gilman and M. Wise, {\it Phys. Rev.}  D20 (1979), 2392.

\item{26.} F.J. Gilman and M.B. Wise, {\it Phys. Rev.}  D21 (1980), 3150.

\item{27.} C.O. Dib, I. Dunietz and F.J. Gilman,
{\it Phys. Rev.}  D39 (1989), 2639; J. Flynn and L. Randall,
{\it Nucl. Phys.} B326 (1989), 31; C.O. Dib, Thesis, Stanford
University, 1990, SLAC--PUB--364 (1990).

\item{28.} G.D. Barr {\it et al.}, {\it Phys. Lett.} B240 (1990), 283.

\item{29.} K.E. Ohl {\it et al.},  {\it Phys. Rev. Lett.} 65 (1990), 1407.

\def\NPB#1#2#3{{\it Nucl. Phys.} B#1 (19#2), #3}
\def\PLB#1#2#3{{\it Phys. Lett.} B#1 (19#2), #3}
\def\PRD#1#2#3{{\it Phys. Rev.} D#1 (19#2), #3}
\def\PRT#1#2#3{{\it Phys. Rep.} #1 (19#2), #3}

\def\TAMU#1{Texas A \& M University preprint CTP-TAMU-#1}

\item{30.} A short talk given by G.T. Park. For more details, see
{\it Phy. Rev.} D48 (1993), 974.

\item{31.} For reviews see: R. Arnowitt and P. Nath, {\it Applied N=1
Supergravity} (World Scientific, Singapore 1983);
H. P. Nilles, \PRT{110}{84}{1}.

\item{32.} \JL, \DVN, and \AZ, \TAMU{68/92}, CERN-TH.6667/92, and
CERN-PPE/92-188.

\item{33.}  CLEO Collab.: E. Thorndike, talk given at the 1993 Meeting of
the American Physical Society, Washington D.C., April 1993.

\item{34.} R. Barbieri and G. Giudice, CERN-TH.6830/93.

\item{35.} S. Bertolini, F. Borzumati, A. Masiero, and G. Ridolfi,
\NPB{353}{91}{591}.

\item{36.} A. Ali and C. Greub, \PLB{293}{92}{226}.

\item{37.} M. Misiak, ZH-TH-19/22 (1992).

\item{38.} \JL, \DVN, and \AZ, \PLB{291}{92}{255}.

\item{39.} \JL, \DVN, and \HP, \PRD{47}{93}{2468}.

\item{40.} R. Arnowitt and P. Nath, \PLB{299}{93}{58} and Erratum;
P. Nath and R. Arnowitt, NUB-TH-3056/92, CTP-TAMU-66/92 (revised).

\item{41.} For a review see, A. B. Lahanas and D. V. Nanopoulos,
\PRT{145}{87}{1}.

\end
\bye